\documentclass[11pt,a4paper]{article}
\usepackage[colorinlistoftodos]{todonotes}

\usepackage{color} 
\PassOptionsToPackage{hyphens}{url}
\usepackage{hyperref}
\hypersetup{colorlinks=true,
	urlcolor=blue,citecolor=blue}
\usepackage{cite}

\usepackage[normalem]{ulem}
\begin{document}

\title{\bf Quassical Computing}
 
\author{Edward H.~``Ned'' Allen\footnote{Chief Scientist, Lockheed Martin, USA.} \and Cristian S. Calude\footnote{Department of Computer Science, University of Auckland, New Zealand.}}

\maketitle

\begin{abstract}
We present a class of hybrid classical systems using quantum 
co-processors and point out that unlike purely quantum computers, such 
hybrids can be both universal and Turing complete; we introduce such 
quantum-classical hybrids as ``quassical."  We discuss the benefits of 
quassical architectures from a theoretical point of view: for some 
classes of problems they achieve computational supremacy. From a 
practical point of view, quassical architectures can also reduce the 
overhead burden imposed by most error correction schemes and minimize 
the challenges of interconnecting qubits in a usefully large connection 
graph. All quantum computing systems are cyber-physical machines and 
thus quassical to at least a trivial degree but only the more profoundly 
quassical hybrids can exhibit an optimum problem-solving capability for 
the amount of quantum resources deployed. Most significantly, quassical 
architectures advance our thinking past that of seeing quantum machines 
as simply quantum embodiments of classical ones and can enliven whole 
new fields of analytical thinking that takes us beyond quantum 
information science per se into a deeper understanding of the duality 
between quantum information and fundamental thermodynamics, possibly 
suggesting unexpectedly useful new technologies. 
\end{abstract}
\thispagestyle{empty}

\section{Quantum vs. classical computing}

Lockheed Martin (LM) pioneered the use of quantum computing in an industrial environment 
in 2010 with the acquisition jointly with the University of Southern 
California of a D-Wave adiabatic quantum computer. So, Professor 
Grandinetti has asked that we speak about the potential of quantum computing 
for our High Performance Computing (HPC) needs. We are glad he did not 
ask us to talk about the actual performance of quantum computing in the 
industrial environment today, because, in spite of a decade's effort by 
a moderately sized team at LM, USC and elsewhere, we are still maturing 
quite juvenile initiatives around the few applications we've identified 
as quantum computing's  ``sweet spots".  Perhaps the main lesson is that 
while quantum computing has advanced tremendously in nearly every 
respect (hardware, software, algorithms, and application concepts), the 
scale of available quantum computers is inadequate for many applications 
and still facing a stiff headwind growing the capability sufficient to 
pose an alternative to massively parallel high performance classical 
systems.

Furthermore, we know that quantum systems cannot do anything classical systems 
cannot also do -- the quantum advantage, if any, is solely one of more 
efficient algorithms  --  sometimes called `quantum speedup' or `quantum 
supremacy'\footnote{C. S. Calude and E. Calude. The Road to Quantum Computational Supremacy, arXiv:\url{1712.01356v2}, March 2018.} and of representation of engineering problems in a more 
natural way. Of course, there is great interest in problems that are so 
intractable that no classical machine could solve with practical 
resources of time and size. Moreover, quantum computation, though 
universal, is significantly restricted in terms of  ``Turing 
completeness"  compared to classical computers. 

The two classical notions 
of universality and Turing completeness are not equivalent --  a fact 
which explains some wrong claims. Turing constructed a universal Turing 
machine capable of simulating any other Turing machine. His result when 
transformed into a definition\footnote{A class of (computing) machines $M=(m_i)$ has a universal machine if there exists a machine $u \in  M$  such that for every $m \in M$  there (effectively) exists an $i$ such that $m(x)=m_i (x)$ and  $u(0^{i}1x) = m_i (x)$, for all bit strings $x$. The input $0^{i}1x$ for $u$ codes the machine index $i$ and the input $x$. Some classes of computing machines have universal machines, others do not. Here are some examples:
\begin{itemize}
\item {\bf Turing theorem.} The class of Turing machines has a universal Turing machine.
\item  {\bf Chaitin theorem.} The class of self-delimiting Turing machines (machines having prefix-free domains) has a universal self-delimiting Turing machine.
\item  {\bf Reversibility theorem.} The class of reversible Turing machines (machines whose computations can be fully undone) has a universal reversible Turing machine. (H. B. Axelsen, R. Gl\"{u}ck.  On reversible Turing machines and their function universality, {\em Acta Informatica} 53 (2016), 509-543.)
\item {\bf Folklore theorem.}  The class of finite state transducers has no universal finite state transducer.
 \end{itemize} }
 reveals that some classes of computing 
machines have universal machines, while others do not. The class of 
quantum gate-based machines\footnote{For details see D. Mermin. Quantum Computer Science, Cambridge University Press, Cambridge, 2007.}, the most studied architecture of quantum computing, 
has \textit{no} universal machine in the sense of Turing, 
as the required equality cannot be satisfied, but a weak form of universality is true when the equality is replaced with an arbitrary close approximation:

\textbf{Solovay-Kitaev theorem}\textit{. Every quantum 
gate operation can be approximated with arbitrary precision by a finite 
sequence of quantum gates from a finite set of quantum gates, like 
$\{$Ising gate and the phase-shift gate$\}$ or the set 
$\{$Hadamard gate, the 
$\pi/8$ gate, the controlled-NOT gate$\}$.}

The second concept derived from Turing's analysis is \textit{completeness.} A class of (computational) machines \textbf{\textit{C 
}}is \textit{Turing complete }if for\textit{ }every Turing machine 
\textit{m }there exists a machine \textit{c} in \textbf{\textit{C 
}}such \textit{that m(x)=c(x),} for all bit strings x. Informally, 
every Turing machine can be (exactly) simulated by some machine in the class 
\textbf{\textit{C}}.

This definition can be used to  show that  some classes of powerful machines have severe restricted computational capacity, which, in particular,
colors our view of the potential future for quantum 
computer: 

\begin{itemize}
\item The class of self-delimiting Turing machines is not Turing 
complete. (Reason: every self-delimiting Turing machine computes only 
strictly partial functions, i.e. functions which are not defined 
everywhere.)
\item The class of reversible Turing machines is not Turing complete. 
(Reason: every reversible Turing machine computes only injective 
functions.)
\item The quantum computing gate model is not Turing complete. (Reason: 
quantum gates compute only total functions, functions defined 
everywhere.)
\end{itemize}

Thus, there are large classes of problems that may not be modeled by injective functions or functions at all, but that map the physical world efficiently and yet are not accessible to quantum computers, or to important classes of classical ones either, but may be amenable to the more elaborate quassical architectures we envision here. The most important such class we suspect is the class of pedagogical problems   --   problems that can be solved with some form of machine learning but are difficult to reduce to functional mathematics otherwise. 

There appear to be two primary obstacles facing those seeking to render quantum computers useful. First, the {\it decoherence}\footnote{That is, the process whereby quantum superposition decays into mutually exclusive classical alternatives, a mixed state, that results in loss of information from a system into the environment and ``robs"  the quantum computer of its power.} {\it problem}\footnote{Symptomatically, a form of quassicality can produce qubits with long coherence times. See M. Stern, G. Catelani, Y. Kubo, C. Grezes, A. Bienfait, D. Vion, D. Esteve, P. Bertet. Flux qubits with long coherence times for hybrid quantum circuits, Phys. Rev. Lett. 113, 123601, September 2014.}    --   when considered as state machines at useful scales, today's quantum models of computing and their corresponding hardware implementations are quite fragile with respect to data integrity; coherence durations are typically but a few milliseconds or micro-seconds and circuit depths are therefore shallow.\footnote{Some results suggest that the Adiabatic Quantum Computing model is more robust against decoherence than the Quantum Gate model. See also B. Tamir, E. Cohen. Notes on Adiabatic Quantum Computers, arXiv:1512.07617, December 2016.} The startlingly unexpected finding over the past two decades that error correction is possible in quantum computing even allowing for the prohibition against copying an unknown quantum state, while rejuvenating enthusiasm for quantum computing, threatens a huge increase in overhead. Microsoft may have the best idea with their topological approach, which encodes information in topologically protected states of qubit arrays via braiding Majorana quasiparticles, thus never allowing the errors to get into the calculation rather than correcting it after they do --  but they are behind the others in demonstrating the achievable validity of their ideas in hardware (though catching up now with encouraging speed). 

Second, the challenge of tying clusters of qubits together with appropriate channels to form mathematically useful gates has turned out to be much more difficult than originally thought   --   perhaps due to the grossly impossible simplifying assumptions one can get away with in mathematics compared to what is realizable in practical and affordable engineering. Like the human brain, quantum computing systems benefit as much, or perhaps more so, from connections as from the multiplicity identity of the qubit. And designing, building and managing communication lines between qubits has turned out to be as hard as or harder than building good qubits. While there are theoretical proposals for quantum architectures using all-to-all connectivity\footnote{W. Lechner, P. Hauke, P. Zoller. A quantum annealing architecture with all-to-all connectivity from local interactions, Science Advance, 1,9 (2015), \url{https://doi.org/10.1126/sciadv.1500838}, S. Puri, C. K.  Andersen, A. L.  and Grimsmo, A.  Blais. Quantum annealing with all-to-all connected nonlinear oscillators, Nature Communications, 8 (2017), \url{http://dx.doi.org/10.1038/ncomms15785}.} so far none only a small portion of a complete graph is offered. We fear that to accommodate larger problems, heroically elaborate networks of quantum channels will be required to significantly improve connectivity and they'll be much slower at communicating than one would desire.\footnote{But one cannot exclude that the connectivity problem may turn out to be yet another practical trade-off, not a fundamental limitation.} So, the first, though tentative, insight we can offer is that HPC is not threatened, at all, by quantum computing   --   at least not yet, and maybe never.

\section{Quantum Number Crunching or Insight?}

Some have said that our difficulties in finding productive applications for the D-Wave machine\footnote{For more information about this approach to quantum computing see C.  McGeoch. Adiabatic Quantum Computation and Quantum Annealing. Theory and Practice, Morgan \& Claypool Publishers, 2014.} were due to the poor quality of the machine, its qubits not having long coherence times, the adiabatic algorithm being too narrow to be useful mathematically, the chimera graph connection scheme of the D-Wave designs being too sparse, and that the machine is not actually making use of quantum effects after all. We have watched and helped as our partners at D-Wave, USC and NASA-Ames have addressed each one of those objections over the years and have been able, more or less, to set each aside  -- not that the D-Wave products cannot be improved, but they are, we've concluded, genuine quantum computers. Moreover, the putative weaknesses of any particular model of quantum computer is a red herring. All flavors of the universal quantum computer, as Feynman believed, will be equally capable at their foundations. The   controversies over which is better may be misguided. While there is some truth to each view, the current arguments do not capture the real weakness or the real benefit of any design.  {\it The valid concerns are and should be how the discovery of the underlying quantum logic and its embodiment in realizable quantum computing hardware is changing and will change our way of thinking about and analyzing truly relevant problems.} This  ``representation'' goal of ours is quite a different end than the search for quantum speedup or quantum supremacy, which may nonetheless come along as a welcome side-effect.

\section{Quantum vs.~digital thinking}

That said, we now turn to what we've found quantum computers can do and do so well that it is hard to see how classical computing, even HPC's, could ever catch up to them in spite of the validity of John Preskill's assertion that  {\it though it may operate according to different physical principles than a classical computer, [a quantum computer] cannot do anything that a classical computer can't do},\footnote{\url{http://www.theory.caltech.edu/~preskill/ph219/ph219_2018}.} -- and most experts would agree with John \dots  as we do.

We start with an analogy: while all number systems can be mapped into the decimal positional one we all use today, non-positional ones, like Roman numerals, are much harder to write algorithms for. This highlights the critical advantages of having the right ``representation"  for any problem: the best representations foster deep insight into how to solve it and moreover, the wrong representation may block any solution whatsoever.  We are increasingly confident that the real value of quantum computing lies not in quantum speedup, or supremacy, but in the profound appropriateness of the quantum insight for guiding us toward solutions to our most important problems. Feynman captured it in his 1982 talk on ``Simulating Physics with Computers"\footnote{Int. J. Theor. Phys. 21 (1982), 467--488.}:  {\it …I'm not happy [he wrote] with all the analyses that go with just the classical theory, because nature isn't classical, dammit, and if you want to make a simulation of nature, you'd better make it quantum mechanical, and by golly it's a wonderful problem, because it doesn't look so easy.}  Feynman never mentions quantum speed-up in his talk because it wasn't so clear at that time that there'd be any -- the fabulous quantum Fourier algorithms of Peter Shor did not appear till more than a decade later. Rather what Feynman was concerned about was the basic inadequacy of classical algorithms in quantum physics, whatever their completeness or precision. His example was that of calculating the probabilities that John Bell's theorizing and Alan Aspect's experiments revealed about the nature of quantum entanglement. Feynman generalized from them:  {\it … the discovery of computers and the thinking about computers has turned out to be extremely useful in many branches of human reasoning. … There are interesting philosophical questions about reasoning, and relationship, observation, and measurement and so on, which computers have stimulated us to think about anew, with new types of thinking. And … computer-type of thinking [when extended to quantum computing] would give us some new ideas …}  At first, one might conclude that Feynman was talking about simple analog computing.  Quantum computers are analog machines after all because they operate analogously to the interactions of quantum particles. But in the years since it has become clear that there's a profound relationship, an intrinsic duality of sorts, between the seemingly continuous-time analogs of physics and the definitively discrete arithmetic of computational analysis when it comes to quantum computing. Because of the quantization of physics, quantum computers have theoretically the precision of digital classical ones\footnote{There are substantial engineering challenges in realizing adequate levels of precision in controlling the quantum computation, a problem potentially as important as decoherence. } and yet behave analogically with respect to quantum physics -- so they are especially suited to simulations of physics -- but their suitability goes well beyond that -- this suggests there is a continuous-discrete duality of sorts to quantum computing. That notion is not well laid out in either computer science or mathematics (though Robinson's {\it non-standard analysis} offers a place to start, a better framework, for it seems to presage the suggestively continuous dualities undergirding the Solovay-Kitaev theorem\footnote{``In the case of quantum computers …The set of possible quantum gates forms a continuum, and it's not necessarily possible to use one gate set to simulate another exactly. Instead, some approximation may be necessary.''  Michael Nielson's blog at \url{http://michaelnielsen.org/blog/the-solovay-kitaev-algorithm}. } -- see for example the Michael Nielsen and Chris Dawson paper on the theorem). Maybe the physicists, or the neurobiologists have an inkling, but no one seems focused on it. 

For quantum computing, {\it digital thinking} -- meaning formulating the problem in terms of classical gates, transforming a classical formulation into a quantum equivalent, and then running the program on a quantum computer -- will not lead to efficient quantum solutions and likely will not see any speed up or representational benefit. Most programs for the D-Wave machine were obtained via this simplistic approach. A better approach, though, is to think from the very beginning of the problem in analogous quantum terms and then naturally solve it on a quantum machine. We have seen this same scenario in the history of molecular computing: reformulating arithmetic operations in terms of molecular operations was abandoned to a direct molecular approach, e.g. using and programming directly a biological transistor and DNA chips. While the initial interest in this field was to tackle NP-hard problems, it was soon realized that they may not be best suited for this type of computation. This insight seems also valid for quantum computing.

\section{Quassical computing}

This quantum {\it analog-discrete duality}, we have concluded, is best exploited for practical applications by conjoining quantum with classical computers in a profoundly intimate way we've called the  ``quassical computer", a term the first author coined\footnote{C. S. Calude, E. Calude, M. J. Dinneen. Adiabatic Quantum Computing Challenges, ACM SIGACT News, 46,1 (2015), 40-61.}. From the very first, at LM we tied the D-Wave machine into our engineering network so that any one of our engineers could call it up from her workstation in MatLab as if it were a MatLab function or script. We did this because our D-Wave, like all QCs being developed today, requires some preliminary classical pre-processing to shape the problem into one the quantum computer can recognize and then to receive the data returned by the D-Wave and shape it into the answer the engineer needs. Many QC offerors will, of course, provide this pre- and post-processing as part of their operating systems so it will be invisible to the user but it's still there of course, visible or not. And, of course, all the quantum computers we've heard of are designed as cyber-physical systems, quantum mechanical systems controlled by digital controllers. So we expect all these new offerings to be quassical in this trivial sense. 

At an intermediate level, a quassical machine might need a fundamentally richer/more expressible language  than a classical programming language (as MatLab) to program it.\footnote{Being stuck with classical programming languages could be an obstacle in using a quassical computer to its full power even if we had one.}  But there is a much more profound sense in which the principle of quassicality can greatly strengthen the quantum computing vision and we will describe that principle now.

Consider a simple Cartesian lattice or matrix. Tracing a path from cell to cell to the right (as in reading a line of text in a Western book) is classical computing, a classical Turing Machine with each cell being a `step' along the tape of a Turing machine, a Turing step. Also going from classical state to classical state implies a measurement and thus any quantum information that might have been present in the cell is collapsed into a single classical state in the ambulatory process. Typically, in real classical computers, such a progression of classical states is not reversible, but that is not fundamentally so, it is so simply because classical computer designers have elected to design them that way.  Thus adding two numbers together to get one cannot be reversed if we lose track of how the sum was originally partitioned. But using techniques invented and perfected by Landauer\footnote{R. Landauer. Irreversibility and heat generation in the computing process, IBM Journal of Research and Development 5 (1961), 183--191.}  and Bennett\footnote{C.~H.~Bennett.  Logical reversibility of computation, IBM Journal of Research and Development 17 (1973), 525--532.}, classical computing can be made fully reversible by keeping a complete historical accounting of how all the partitions are collapsed. And so, following the notion that irreversible computing is merely a special case of reversible computing where the accounting is ignored or discarded (uses less memory of course and speeds things up as complete accounting is intensive work), we elect to do our classical computing reversibly so that one may read a line of classical argument either right to left or left to right --  doing a calculation or undoing it precisely. This decision is not so arbitrary either, for if information is physical, as most have argued since the seminal papers of Landauer and Bennet, discarding complexity must at least partially collapse phase space and thus increase entropy and hence incur an energy cost. So, it necessarily generates heat whereas reversible computing is or can be fully adiabatic.   
 
We define traveling downward (upward) from one line of cells to the line below (above) in our Cartesian lattice as a unitary quantum evolution which is always reversible, so no `measurement' takes place when reading downward (upward) through the cells. Accordingly, any cell in the lattice may be interpreted (reinterpreted) as a quantum or a classical datum: we can reinterpret the classical state (described by some vector) as a quantum state (typically a superposition of some of quantum states, but also described by a vector); reinterpretation takes place wholly in the mind --  there is no experimental counterpart, so there is no wave function collapse into classical information. We can do this because classical information is merely a special case of quantum information, a projection of quantum information onto a less complex thermodynamic space. Both are vectors existing in the same vast Hilbert space but the classical vector is merely an irreversible (and hence exothermic) compression in complexity of the quantum one, a lower dimensional slice through Hilbert space. Heat emission is intrinsic to quantum measurement processes; when measured data is funneled down in complexity into whatever the measuring instrument can handle (within the quantum rules, e.g., the uncertainty principles, the no-cloning theorem, the Holevo bound, etc.), entropy is increased, heat is generated.  But in our quassical model, the additional data is not lost just stored elsewhere so the process can be adiabatic.\footnote{M. P. Frank. Back to the Future: The Case for Reversible Computing, \url{arXiv:1803.02789v2}, March, 2018.}

\section{Trajectory length and slope through the quassical cube}

Using this Cartesian lattice as a sketch pad, one can draw trajectories for computation using hybrid quantum-classical circuits, that is, quassical circuits. And most significantly, one can measure the distance between states that can be reached by combinations of traversing between states quantumly (upward or downward in direction) and classically (sideways in direction) --  contracting the complexity  from quantum to classical, and then re-expanding it back to quantum as required --  no information is lost, no heat rejected. In order to assure reversibility, of course, one must always keep the information put aside each time a step is taken to the right and reabsorb that information each time a step is taken to the left. Likewise, each time a qubit is measured, collapsing its information content down to a classical bit, the information not embedded in the classical bit must be accounted for and stored off-line so it can be restored in the reverse operation --  reinterpreting a bit as a qubit. This is an imagined reversal of the quantum measurement process. `Off-line' here means in a cell or cells not part of the two dimensional pathway through the lattice --  thus implying the existence of a third dimension (which, of course, may be merely a remote and unused portion elsewhere on the same lattice). And the stored data must be stored in the correct order according to the way it was generated as is required to achieve the reversal (or tagged so that the correct sequence can be reconstructed when re-expanding it to its unreconstructed status). Thus, the complete embodiment of the idealized quassical computer is a three dimensional volume, a cubical information structure each slice of which contains quassical information. A `circuit' in this embodiment is any pathway from cell to adjacent cell to adjacent cell, etc., traced through the quassical cube, making use of an orthogonal dimension storehouse as necessary. Consequently, the third dimension, meaning the order, structure, and, thus, entropy of the storehouse is determined by the history of calculation because it must contain all data required to render the calculation reversible --  so it is not unconstrained and thus not a true  ``degree of freedom"  in the strict sense of the phrase.

Now we define the length of any pathway through the quassical cube, for in that definition lies one of the fundamental aspects of the quassical insight. It is related to the circuit depth:  ``the depth of a circuit''  through a computer, as Preskill writes,  {\em is the number of time-steps required, assuming that gates acting on distinct bits can operate simultaneously (that is, the depth is the maximum length of a directed path from the input to the output of the circuit). The {\em ``width of a circuit''}, the maximum number of gates (including identity gates acting on  ``resting"  bits) that act in any one time step, quantifies the storage space used to execute the computation.}\footnote{In a very nice and clear lecture on topological quantum computing delivered by Microsoft Station Q scientist Dr R.~Lutchyn (\url{https://goo.gl/Nw5mRH}) it is suggested that an attractive quassical architecture might use ``conventional quantum computing circuits"  to perform calculations while topological qubits would be employed to store quantum information. Thus a physical realization of a profoundly quassical system might be one in which the topological qubits form a new ``third"  dimension above the ``conventional qubits"  in our quassical cube.} We can take a 2-dimensional slice through our quassical cube of cells and each cell of that slice is a special gate, a quassical gate (one that exhibits quantum-classical duality). 

There are some obvious things to say about a quassical cube: first, it can contain all meaningful data\footnote{The meaning of  ``meaningful'' here stems from the physics of the process which must remain fully adiabatic in the sense that all information is conserved and IAW Landauer, i.e., just as energy is conserved, so must be information.}  and thus all physical states (implied to include any associated data, their `quassical meanings') have a home somewhere in the cubic volume. Second, each datum in a quassical cube has the dual quantum-classical character (so each unit  ``cell"  is really at least a column of cells storing both the state itself and data above and below that cell that is required to traverse it in any of the four possible directions: up, down, right, left (recall this  ``additional dimension"  is not independent) so each is a  ``slice"  through a qubit. It may, of course, be better understood if each cell is considered a quassical unit computer, a universal quassical gate, containing the minimum functionality to serve both its local and its system-level purpose. 

Is the model of quassicality described above equivalent with adiabatic quantum computing?\footnote{C. McGeoch. Personal communication to C. Calude, April 24, 2018.}  The answer is negative: quassical computing is more powerful than quantum annealing/adiabatic quantum computing because it is Turing complete, whereas the latter is not by a well-known equivalence\footnote{D. Aharonov, W. van Dam, J. Kempe, Z. Landau, S. Lloyd, O. Regev. Adiabatic quantum computation is equivalent to standard quantum computation, SIAM J. Comput., 37(1) (2007), 166--194.}.

Could  the model of quassicality described above  lead  to quantum  supremacy?\footnote{K. Svozil. Personal communication to C. Calude, May 3 , 2018.}  A quantum computational supremacy experiment has to prove both a lower bound and an upper bound. The upper bound comes from 
from the running time of a quassical algorithm and the
 lower bound is 
necessary for proving that no classical computer can match it. Proving  lower bounds is notoriously  more difficult to prove than  upper bounds; verifying them experimentally is even more demanding$^1$.

One of the first  non-trivial examples of  quassical algorithms is related to Grover's quantum algorithm (which, we recall,  solves the following problem: access to an unsorted quantum database that can be queried with a quantum input is given and asked if it contains a specific entry). Grover's algorithm offers a {\em provable speedup}, though not an exponential one and, more importantly, the problem it solves is far from being realistic: the cost of constructing the quantum database could negate any advantage of the algorithm, and in many classical scenarios one could do much better by simply creating (and maintaining) an ordered database.  In 2005 Lanzagorta and Uhlmann\footnote{M. Lanzagorta, J. K.  Uhlmann. Hybrid quantum-classical computing with applications to computer graphics,
ACM SIGGRAPH 2005 Courses, 2005, \url{http://doi.acm.org/10.1145/1198555.1198723}.}  used Grover's algorithm as a quantum subroutine of a classical algorithm for solving problems in image processing. This quassical approach is provably more efficient than the direct use of Grover's algorithm because the cost of preparing the quantum  ``database"  can be spread out over several calls.

 Abbott at al.\footnote{A. A. Abbott, C. S. Calude, M. J. Dinneen, R. Hua. A Hybrid Quantum-Classical Paradigm to Mitigate Embedding Costs in Quantum Annealing, \url{arXiv: 1803.04340}, March 2018. } describe a quassical algorithm for quantum annealers that mitigates the need to embed problem instances within the (often highly restricted) connectivity graph of the annealer. More precisely, the paper shows how a raw speedup that is negated by the embedding time can nonetheless be exploited to give a practical speedup in solving certain computational problems, like the maximum weight independent set problem.  When applied to a large enough batch of instances of such a problem, the quassical algorithm theoretically outperforms any classical algorithm solving the problem.  While an experimental in-depth  study on the D-Wave 2X machine of such a problem was not able to confirm a quantum speedup, the advantage of the hybrid approach was robustly verified.

Now imagine that we are able to refine our theory to the point where each quassical cell of the cube is made so finely that the process of stepping from one cell to the next can be considered a continuous one defining a pathway that is a continuous process i.e., drawn as a continuous line --  not a discrete walk of many steps. Then it becomes clear that, in general, the shortest distance between any two locations in the cube (between, that is, an input problem statement and its answer output) must be along a diagonal, cutting across quantum and classical evolutions alike and thus must include both quantum-like and classical-like operations, for it is only in rare and
 improbable cases that a purely classical route or a purely quantum route will be optimal.
  In a perfect world this shortest distance is akin to the algorithmic complexity of the problem (as proposed by Chaitin and Kolmogorov). This, then is the first fundamental aspect of the quassical architecture: it realizes the least circuit depth, the least complex pathway for executing  calculation.

\section{The Jozsa conjecture}

A conjecture posed by Richard Jozsa, noted that measurement-based models of quantum computing\footnote{R. Jozsa (An introduction to measurement based quantum computation, \url{arXiv:quant-ph/0508124}, August 2015) gives two examples of measurement based QC:  ``one way quantum computing'', and  ``teleportation quantum computing".  See also Jacob Miller, Stephen Sanders, Akimasa Miyake. Quantum supremacy in constant-time measurement-based computation:  A unified architecture for sampling and verification, Phys. Rev. A 96, 062320 (2017).}   {\it …provide a natural formalism for separating a quantum algorithm into classical parts and quantum parts … In [the measurement based] formalism any quantum computation is viewed as a sequence of classical and quantum layers. The total quantum state is passed from one quantum layer to the next [this is the vertical traverse in our quassical lattice] and the quantum actions carried out in the next layer are determined by classical computations on measurement outcomes from previous layers}  [this is the horizontal traverse across our lattice]. He goes on then to his important conjecture that   {\it any polynomial time quantum algorithm can be implemented with only O$(\log n)$ quantum layers interspersed with polynomial time classical computations. This conjecture, asserting an exponential reduction in the essential  ``quantum content"  of any quantum algorithm, has no analogue in classical complexity theory … Intuitively we are conjecturing that polynomial time classical computation needs relatively little  ``quantum assistance"  to achieve the full power of polynomial time quantum computation.}  We note that all quantum computing models can be rewritten or reinterpreted as measurement based systems, so his conjecture has wider implications than his paper asserts. 

Applying his conjecture to our quassical lattice implies that the full power realizable from quantum computing (that is, the vertical distance from an input point in the quassical cube to an output point that is precisely below or precisely above it) can be achieved by a diagonal line that also traverses many classical states. Diagonals will have more depth (more time steps) than either vertical or horizontal lines but they also have two critical features that render them especially appealing: they greatly extend the utility of circuits fashioned from qubits with limited coherence time, and they substantially reduce the difficulties of achieving more complete connection graphs.  

The `coherence time' advantage stems from the potential for limiting the amount of time the computer's state must be represented by purely quantum data. All manifestations of the qubit (ion traps, superconducting circuits, polarized photons, even topological braids of Majorana fermions, etc.) suffer decoherence to some extent and require error correction in the form of additional overhead to achieve acceptable levels of fault tolerance.\footnote{This is not as acute in the quantum annealing paradigm, T. Abash, D. A. Lidar. Decoherence in adiabatic quantum computation, \url{arXiv:1503.08767}, June 2015.}
These error correction circuits pose what can become an unbearable burden by increasing the number of physical qubits and qubit-to-qubit communication channels required to form a single, fault-tolerant  ``logical qubit"  (tens, hundreds, or even thousands of physical qubits and connections per logical qubit up to a limit posed by the threshold theorem). The quassical architecture would, in principle, allow the design of pathways through the system that shortened the amount of time the information must remain in its quantum state and that could maintain the state of the computer in classical vectors transmuting them back into quantum vectors only for the purpose of performing a portion of the calculation that is best done in the Hilbert space of quantum computing (e.g., the quantum Fourier transform, quantum amplification, etc.). The coherence time through a quassical lattice is proportional to the slope of the diagonal --  so that reducing the amount of time the computer's executive spends in quantum space necessarily entails spending more time in classical space and rotating the diagonal trajectory toward the classical limit and implying more time steps. This may or may not be advantageous from a quantum speedup point of view, but it is certainly advantageous from a decoherence point of view.  

Another advantage of the quassical architecture relates to interconnections between qubits. The seriousness of the interconnection challenge has been truly recognized only recently as more engineering groups are attempting to create useful systems of many qubits. In the D-Wave designs, all the qubits are arrayed on a chip, a 2-dimensional grid in Chimera graph, which is almost, but not a planar graph. The connections are only between nearest neighbors, so that most of the qubits on a pathway between any two qubits we may want to connect must be sacrificed to fashion a quantum connection channel. A little thought about graph theory and topology reveals that this problem is intrinsic to more or less every quantum computer design one can engineer with a non-complete graph architecture in our three dimensional world. We refer to this as the  ``graph"  or  ``topological"  constraint inherent in the architecture of quantum computers. Classical computers of course have the same problem, but it is much less acute because of the absence of quantum fragility and the facility with which classical bits may be copied and amplified when routing them through arbitrary networks of communication channels. The quassical architecture can, in principle, move some of the information through classical channels greatly mitigating the topological graph problem. 

One of the more natural applications of the quassical architecture became evident to the first author in the course of an experiment  conducted to use the D-Wave machine to a deep learning problem. 
The experiment\footnote{S. H. Adachi, M. P. Henderson. 
Application of Quantum Annealing to Training of Deep Neural Networks,
\url{arXiv:1510.06356}, October 2015.} 
studied a class of pedagogical methods based on using the D-Wave machine as the `instructor' for a deep learning network
and, as many have since discovered, quantum computers can be used to train neural networks more efficiently (less circuit depth) and with more accuracy and precision per training cycle than using classical techniques. Pedagogy, wherein the  ``professor"  is quantum and the student is a classical network, is a natural and suggestive application of quassical computing. In particular, one would like to know whether a simple pedagogical quassical gate could be designed and how  ``universal"  that gate would be. Here universal is intended in the sense defined above for the quantum gate model of quantum computing.   

\section{Conclusion: the future glimpsed from our observation point}

We hope we have conveyed our intuitions that quantum computers will not make classical machines obsolete: quite the contrary, they will be integrated into classical machines as co-processors of a sort perhaps like the ways Jozsa sets forth in his essay on measurement-based quantum computing. So, we hold that quantum computing will extend and enhance classical computing --  not supplant it. Moreover, the best integrations of quantum| and classical computers\footnote{Reversible  classical computing, see the Reversibility Theorem above,  offers probably the most natural candidate.}  will exploit non-trivial, profound quassicality because it offers a more cost-effective pathway to the full power of quantum computing without excessive error correction overhead or the achievement of exotic coherence times for qubits --  even crumby qubits can excel in quassical architectures.\footnote{Measurement-based models of quantum computing,  discussed above, could be
viewed as a different form of  ``quassical computing''  wherein the quantum computer, like in  the neural network training algorithm,  is used
as a co-processor within a larger iterative classical algorithm.}
 In addition, the topological connectivity graph constraint may be largely lifted for quassical architectures, providing all the more appeal for quassical designs. But perhaps the most important legacy of quassical architectures will be that they advance our thinking past that of seeing quantum machines as simply quantum embodiments of classical algorithms and machines. They will enable a whole new field of quassical thinking that extends beyond quantum information science to support a new and clearer understanding of quantum information science and its fundamental thermodynamic dual and, thus, spawn whole families of unanticipated new technology.   

\section*{Acknowledgements}
 We thank Dr.~Catherine McGeoch and Prof. Karl Svozil  for enlightening discussions and critical comments, particularly  about universality and completeness. We also thank Dr. Steven Adachi for suggestions which improved the presentation. Calude was supported in part by the Quantum Computing Research Initiatives at Lockheed Martin. 
 
\end{document}